%% file: sample-sigconf.tex
\documentclass[sigconf]{acmart}
\usepackage{hyperref}
\usepackage{breakurl}
\usepackage{url}
\usepackage{balance}
\usepackage{booktabs} % For formal tables

\usepackage{bbm}

\usepackage{amsmath}
\usepackage{amsthm}
\usepackage{enumitem}
\usepackage{amsfonts}
\usepackage{amssymb}
\usepackage{microtype}
\usepackage{tabularx}
\usepackage{siunitx}

\usepackage{multirow} 

\usepackage{graphicx} 

\usepackage[ruled]{algorithm2e} % For algorithms

\setcopyright{rightsretained}
\usepackage{graphicx}
\usepackage{subcaption}

\begin{document}

\fancyhead{}

\newcolumntype{L}[1]{>{\raggedright\arraybackslash}p{#1}}
\newcolumntype{C}[1]{>{\centering\arraybackslash}p{#1}}
\newcolumntype{R}[1]{>{\raggedleft\arraybackslash}p{#1}}

\title{Bid Optimization by Multivariable Control in Display Advertising}
%\titlenote{Produces the permission block, and
%  copyright information}
\copyrightyear{2019} 
\acmYear{2019} 
\setcopyright{acmlicensed}
\acmConference[KDD '19]{The 25th ACM SIGKDD Conference on Knowledge Discovery and Data Mining}{August 4--8, 2019}{Anchorage, AK, USA}
\acmBooktitle{The 25th ACM SIGKDD Conference on Knowledge Discovery and Data Mining (KDD '19), August 4--8, 2019, Anchorage, AK, USA}
\acmPrice{15.00}
\acmDOI{10.1145/3292500.3330681}
\acmISBN{978-1-4503-6201-6/19/08}

%\subtitle{Extended Abstract}
%\subtitlenote{The full version of the author's guide is available as
%  \texttt{acmart.pdf} document}
\author{Xun Yang, Yasong Li, Hao Wang, Di Wu, Qing Tan, Jian Xu, Kun Gai}
%\authornote{di.wudi@alibaba-inc.com}
 \affiliation{%
   \institution{Alibaba Group}
   \city{Beijing}
   \country{P.R.China}
 }
 \email{{vincent.yx,yasong.lys,wh111044,di.wudi,qing.tan,xiyu.xj,jingshi.gk}@alibaba-inc.com}

%\authornote{Dr.~Trovato insisted his name be first.}
%\affiliation{%
%  \institution{Institute for Clarity in Documentation}
%  \streetaddress{P.O. Box 1212}
%\authornote{The secretary disavows any knowledge of this author's actions.}
%\affiliation{%
%  \institution{Institute for Clarity in Documentation}
%  \streetaddress{P.O. Box 1212}
%\email{webmaster@marysville-ohio.com}
%
%\author{Lars Th{\o}rv{\"a}ld}
%\authornote{This author is the
%  one who did all the really hard work.}
%\affiliation{%
%  \institution{The Th{\o}rv{\"a}ld Group}
%  \streetaddress{1 Th{\o}rv{\"a}ld Circle}
%\email{larst@affiliation.org}

\renewcommand{\shortauthors}{X. Yang et al.}

\begin{abstract}
Real-Time Bidding (RTB) is an important paradigm in display advertising, where advertisers utilize extended information and algorithms served by Demand Side Platforms (DSPs) to improve advertising performance. A common problem for DSPs is to help advertisers gain as much value as possible with budget constraints. However, advertisers would routinely add certain key performance indicator (KPI) constraints that the advertising campaign must meet due to practical reasons. In this paper, we study the common case where advertisers aim to maximize the quantity of conversions, and set cost-per-click (CPC) as a KPI constraint. We convert such a problem into a linear programming problem and leverage the primal-dual method to derive the optimal bidding strategy. To address the applicability issue, we propose a feedback control-based solution and devise the multivariable control system. The empirical study based on real-word data from Taobao.com verifies the effectiveness and superiority of our approach compared with the state of the art in the industry practices.
\end{abstract}

\begin{CCSXML}
<ccs2012>
<concept>
<concept_id>10002951.10003227.10003447</concept_id>
<concept_desc>Information systems~Computational advertising</concept_desc>
<concept_significance>500</concept_significance>
</concept>
<concept>
<concept_id>10002951.10003260.10003272.10003275</concept_id>
<concept_desc>Information systems~Display advertising</concept_desc>
<concept_significance>500</concept_significance>
</concept>
</ccs2012>
\end{CCSXML}

\ccsdesc[500]{Information systems~Computational advertising}
\ccsdesc[500]{Information systems~Display advertising}

\keywords{Real-Time Bidding; Bid Optimization; Feedback Control; Display Advertising}

\maketitle

\input{samplebody-conf}
\balance
\bibliographystyle{ACM-Reference-Format}
\bibliography{acmart}

\end{document}

%% file: samplebody-conf.tex
\section{Introduction}
		
	Online display advertising has been an increasely significant business. According to the Internet Advertising Bureau (IAB) report, internet advertising revenues for the full year of 2017 increased 21.4\% over 2016 and totaled \$88.0 billion in the United States, with display advertising\footnote{Display-related ad formats include: Banner and Video} accounting for approximately \$39.4 billion \cite{iab-hy-2017}. In online display advertising, advertisers pay a certain price for the ad opportunity to show their ads. Real-Time Bidding (RTB)\cite{wang2015real} is the most popular paradigm in display advertising. RTB enables advertisers to bid for the ad opportunity at the impression level, and the bidder with the highest price wins the opportunity to show its ad. Specifically, the bid price for each ad opportunity could be individually different based on its utility and cost, which allows advertisers to leverage extended information and algorithms served by DSPs\cite{yuan2013real}. %Generally, the value from advertising could be quantified by the generated conversions, which is what most advertisers ultimately want\cite{kitts2017ad}
	
	A common problem for DSPs is to help advertisers gain as much value as possible within the budget\cite{wu2018budget}. There has been some bidding strategies and algorithms\cite{wu2018budget, cai2017real} proposed to maximize the value from advertising with the budget constraint. Therefore, advertisers could simply set the budget of the campaign, and DSPs would calculate the bid price on behalf of advertisers. 
	
	However, apart from the budget constraint, advertisers would routinely add certain key performance indicator (KPI) constraints that the advertising campaign must meet\cite{kitts2017ad}. Advertisers set such KPI constraints because the campaign with a single budget constraint may suffer from huge changes of the traffic caused by the volatilities of the bidding environment. For example, the ad opportunities of certain days could become so expensive that the advertiser cannot afford to spend all budget on them. One solution is constantly adjusting the daily budget to control the investment, which is costly and even impractical for advertisers. Another solution is to set certain KPI constraints. Some KPI constraints, such as the cost-per-mille (CPM) and cost-per-click (CPC) constraints, have strong influence on the total cost of a campaign. By setting such KPI constraints, advertisers could cast a restriction on the total cost and avoid spending all budget when the ad opportunities are not worthwhile, which frees advertisers from heavy labor on frequently adjusting campaign settings. Furthermore, KPI constraints also serve as a real-time proxy to regulate the advertising performance. In most cases, advertisers ultimately want conversions. However, the conversion is sparse and delayed, which prohibits advertisers to evaluate the advertising performance in real time. As a result, advertisers use the KPI exposed by DSPs to evaluate the expected value from advertising and set it as a constraint to ensure that the performance of advertising is under control. 
	%However, apart from the budget constraint, advertisers would routinely add certain key performance indicator (KPI) constraints that the advertising campaign must meet\cite{kitts2017ad}. Advertisers set such KPI constraints to reduce the risk that comes along with the volatilities of the bidding environment. Since the bidding environment may fluctuate fiercely across days\cite{zhang2016feedback}, advertisers have to daily adjust the budget to avoid risks caused by huge changes of the traffic. For example, the ad opportunities could become so expensive that the advertiser cannot afford to spend too much. However, constantly adjusting budget is costly and even impractical for advertisers. By setting a KPI constraint, advertisers could cast a restriction on the total cost and indirectly control the investment. Secondly, the KPI constraint serves as a real-time proxy to regulate the advertising performance. In most cases, advertisers ultimately want conversions. However, the conversion is sparse and delayed, which prohibits advertisers to evaluate the advertising performance in real-time. As a result, advertisers usually use the KPI to evaluate the expected yield from advertising and set it as a constraint to ensure the advertising performance is acceptable. 

	In this paper, we focus on the CPC constraint, which is one of the most common KPI constraints\cite{zhang2014optimal, zhang2016feedback}. We propose the optimal bidding strategy that maximizes the quantity of conversions from advertising under the budget and CPC constraints. In this work, bid optimization is formulated as a linear programming problem, and the primal-dual method is leveraged to derive the optimal bidding strategy. Our methodology could be generalised to other cost-related KPI constraints such as the CPM constraint.
	
	Furthermore, we propose the multivariable control system based on the optimal bidding strategy to address the applicability issue, particularly the dynamic environment, as we apply the bidding strategy in the industrial situation. Based on the analysis of the hyper parameters in the bidding strategy, we claim that the hyper parameters have strong control capabilities on achieving corresponding constraints, and devise the independent PID control system. Taking into consideration the coupling effect, we further improve the performance of the system by proposing the model predictive control system.
	
	Moreover, the proposed systems are implemented and evaluated on real industrial datasets. Experiments on the real datasets from Taobao.com show that the systems have strong control capability on achieving the constraints. We also compare our approach with the state of the art in the industry practices, and the result reveals the superiority of our method. The main contributions of our work can be summarized as follows: 
	
\begin{enumerate}[align=right,leftmargin=0.17in]
\item We propose the optimal bidding strategy that maximizes the quantity of conversions under the budget and CPC constraints.
\item We devise the multivariable control system to deal with the dynamic environment when applying the bidding strategy in the industrial situation. 
\item Extensive experiments are conducted and the results demonstrate the advantage of our approach.
\end{enumerate}

The rest of this paper is organized as follows: we formulate the problem and derive the optimal bidding strategy in Section 2. Section 3 addresses the applicability issue and presents the multivariable control system. Experiments and evaluations are conducted in Section 4, followed by the related work in Section 5. We conclude our work in Section 6.

\section{Bidding Strategy}

	In this section, we firstly review some preliminary knowledge of the RTB eco-system, and afterwards formulate the bid optimization problem. Then we derive the optimal bidding strategy, and take a discussion on the characteristics of the bidding strategy.

\subsection{RTB Eco-system}

\begin{figure} 
  \centering
    \includegraphics[width=0.48\textwidth]{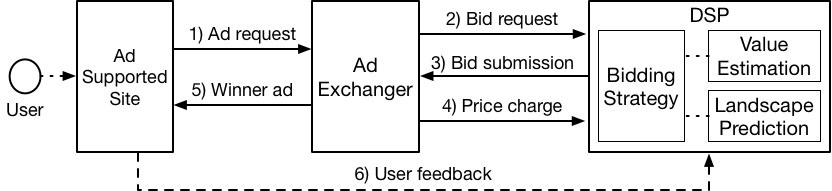}
  \caption{RTB Eco-system}
  \label{fg:RTB-system}
\end{figure}

To make this paper self-contained, we briefly introduce RTB eco-system and its related techniques. The work flow of RTB is illustrated as Fig. \ref{fg:RTB-system}, and each step is as follows: 1) A user visits an ad-supported site, and the site sends an ad request to the ad exchanger. 2) The ad exchanger initiates an auction and requests bids from DSPs. 3) DSPs submit the bid price along with the ad to the ad exchanger on behalf of advertisers. 4) The ad exchanger holds the auction and charges the winner DSP for the ad opportunity. 5)The winner's ad is sent to the site. 6) User feedback afterwards would be sent back to the corresponding DSP. To avoid redundancy, we focus on step 3), 4) and 6), which are highly related to our work.

%预估技术，ctr和cvr
A DSP calculates a bid price for the advertiser by its bidding strategy when it receives a bid request from the ad exchanger. Since conversions are the target event of most advertisers, almost all bidding strategies (including ours) rely heavily on the ability of learned models to estimate ad click-through rate (CTR) \cite{mcmahan2013ad} and conversion rate\footnote{The conversion rate is conditioned on the click} (CVR) \cite{lee2012estimating}. In addition, some DSPs may also base their bidding strategies on the prediction of the winning price (bid landscape prediction). CTR/CVR estimation and landscape prediction by themselves are heavily studied problems \cite{zhou2018deep, yang2017bayesian, wang2016functional}, which is beyond our scope. Therefore, we just assume the estimation and prediction problems have been solved, and the expected probability of the click and conversion could be quantified by CTR and CVR respectively. %本身是一个很大问题，我们这里不展开讨论，我们的bidding策略也需要ctr和cvr，我们默认已经有了。

When a DSP wins the opportunity to show the ad in an auction, it is charged a price. The price is equal to the second highest bid price under the generalized second price (GSP) auction mechanism \cite{edelman2007internet}, which is widely adopted in industrial platforms. There are also some other auction mechanisms such as Vickrey-Clarke-Groves auction mecahnism (VCG) \cite{nisan2007algorithmic}. In this paper, without loss of generality, we base the discussion and formulation on the most common GSP auction mechanism.

%反馈：点击和转化，在现实中我们要根据反馈来调控
Any feedback on the ad from the user, such as clicks and conversions, would be sent back to the corresponding DSP. The DSP could utilize such feedbacks to train the prediction models, and timely adjust its bidding strategy. In addition, such feedbacks would be integrated and exposed to advertisers by the DSP. In our proposed system, we take advantage of such feedbacks and continuously fine tune the bidding strategy across the lifetime of the advertising campaign.

\subsection{Problem Formulation}

Suppose there are $N$ ad opportunities in a day, and we index each ad opportunity by their generated order as $opportunity_i$. Each ad opportunity has different value for the advertiser, and we use $v_i$ to represent the value of $opportunity_i$ for the advertiser. Based on $v_i$, the bid price $bid_i$ is calculated and submitted to the ad exchanger. Each ad opportunity has a winning price $wp_i$. From the perspective of the advertiser, $wp_i$ equals to the highest bid price of other advertisers. If $bid_i$ is higher than $wp_i$, which means that the advertiser would win the ad opportunity and be charged $wp_i$ under the GSP auction mechanism, we set $x_i$ to be 1, and 0 otherwise. The total $Value$ and $Cost$ of the advertising campaign is formulated as Eq. \eqref{ad_value} and Eq. \eqref{ad_cost}.  
\begin{equation} \label{ad_value}
\begin{aligned}
Value = \displaystyle\sum\limits_{i=1...N}x_i \cdot v_i \ \ \\
\end{aligned}
\end{equation}

\begin{equation} \label{ad_cost}
\begin{aligned}
Cost = \displaystyle\sum\limits_{i=1...N}x_i \cdot wp_i \ \ \\
\end{aligned}
\end{equation}

We formulate CPC in Eq. \eqref{eq_cpc}. It is worth noting that we replace the real click with CTR, which provides us with a more concise formulation. Such a replacement could largely facilitate our theoretical analysis, and has trivial influence in the following practical system design.
\begin{equation} \label{eq_cpc}
\begin{aligned}
CPC = \frac{\displaystyle\sum\limits_{i=1...N}x_i \cdot wp_i}{\displaystyle\sum\limits_{i=1...N}x_i \cdot CTR_i} \ \ \\
\end{aligned}
\end{equation}

The conversion is what advertisers ultimately want. Therefore, we quantify $v_i$ by $CTR_i \cdot CVR_i$. Please note that $CTR_i$ must be considered since conversions could only be generated after a click and CVR is conditioned on a click. We summarize the problem as follows and formulate it as \eqref{lp1}.
\begin{itemize}[align=right,leftmargin=0.17in]
\item We maximize the quantity of conversions with the budget $B$, and guarantee that CPC does not exceed a given value $C$.

\begin{align}
			 &\underset{x_i}{\textup{max}} & & \displaystyle\sum\limits_{i=1...N}x_i \cdot CTR_i \cdot CVR_i \tag{LP1}\label{lp1} \\
			&\textup{s.t.} & &  \displaystyle\sum\limits_{i=1...N}x_i \cdot wp_i \leq B \label{con_p} \\
		&	& &  \frac{\displaystyle\sum\limits_{i=1...N}x_i \cdot wp_i}{\displaystyle\sum\limits_{i=1...N}x_i \cdot CTR_i} \leq C \label{con_q} \\
		&\textup{where} & &  0 \leq x_i \leq 1, \forall i \nonumber 
\end{align}
\end{itemize}

\subsection{Optimal Bidding Strategy}
The problem \eqref{lp1} is actually a linear programming problem\cite{schrijver1998theory}, which is to find the optimal $x_i$ to maximize the target function with the linear constraints. There has been many algorithms proposed to directly solve such a problem, however, we aim to derive the optimal bidding strategy instead of the allocation strategy. In other words, we do not essentially care about the value of $x_i$, but the underlying bidding strategy that intrinsically affects $x_i$. With such a consideration, we creatively resort to the primal-dual method. Every linear programming problem, referred to as a primal problem, can be converted into a dual problem\cite{dantzig1983reminiscences}. In addition, the optimal primal solution can be obtained by the corresponding dual solution according to the duality theorem\cite{boyd2004convex}. Such mathematics characteristics shell some insights on us, and we integrate the primal space and the dual space to derive the following theorem:
\begin{theorem}\label{op_therm}
The 	optimal bidding strategy formulation is: 
\begin{equation} \label{eq_opt_func}
\begin{aligned}
bid_i =
\frac{1}{p + q} \cdot CTR_i \cdot CVR_i + \frac{q}{p + q} \cdot CTR_i \cdot C
\end{aligned}
\end{equation}
\end{theorem}

The optimal bidding strategy is stated in Eq. \ref{eq_opt_func}, where $p$ and $q$ are hyper parameters incorporated from the dual space, and correspond to the optimal dual solution. We will investigate the properties of $p$ and $q$ in later sections. For now, we give the proof for Thm. \ref{op_therm}.

\begin{proof}
\eqref{lp1} is converted to the dual problem:
\begin{align}
			 &\underset{p, q, r_i}{\textup{min}} & & B \cdot p + \displaystyle\sum\limits_{i=1...N}r_i\tag{LP2}\label{lp2} \\
			&\textup{s.t.} & &  wp_i \cdot p + (wp_i - CTR_i \cdot C)q + r_i \geq v_i , \forall i \label{lp2st1} \\
		&\textup{where}	& &  p \geq 0, \nonumber \nonumber \\
		&				& &  q \geq 0, \nonumber \\
		&				& &  r_i \geq 0 , \forall i \nonumber \\
		&				& &  v_i = CTR_i \cdot CVR_i , \forall i \nonumber 
\end{align}

Assume the optimal solution for the primal problem \eqref{lp1} is ${x_i^*}, \forall i=1,...,n$, and the optimal solution of the corresponding dual problem \eqref{lp2} is $p^*$, $q^*$, $\{r_i^*|i=1,...,n\}$. According the theorem of complementary slackness, we obtain:
\begin{equation} \label{eq_slackness}
\begin{aligned}
x_i^* \cdot (v_i - wp_i \cdot p - (wp_i - CTR_i \cdot C)q - r_i) =0 , \forall i \ \ \\
\end{aligned}
\end{equation}
\begin{equation} \label{eq2}
\begin{aligned}
(x_i^* - 1) \cdot r_i^* = 0 , \forall i \ \ \\
\end{aligned}
\end{equation}

We delicately set $bid_i^* = \frac{1}{p^* + q^*}{CTR_i \cdot CVR_i} + \frac{q^*}{p^* + q^*}{C \cdot CTR_i}$, then we transform Eq. \ref{eq_slackness} to Eq. \ref{eq1_bid_slackness}:
\begin{equation} \label{eq1_bid_slackness}
\begin{aligned}
x_i^* \cdot ((bid_i^* - wp_i)(p^* + q^*) - r_i^*) =0 , \forall i \ \ \\
\end{aligned}
\end{equation}

\begin{itemize}[align=right,leftmargin=0.17in]

\item According to Eq. \eqref{eq1_bid_slackness}, if the campaign wins $opportunity_i$, i.e. $x_i^* > 0$, then $(bid_i^* - wp_i)(p^* + q^*) - r_i^* = 0$. Meanwhile, $p^* \geq 0, q^* \geq 0, r_k \geq 0$, so $bid_i^* \geq wp_i$.
\item If the campaigns loses $opportunity_i$, i.e. $x_i^*=0$, we could deduce from Eq. \eqref{eq2} that $r_i^* = 0$. Therefore, according to Eq. \eqref{lp2st1}, we obtain $(bid_i^* - wp_i)(p^* + q^*) \leq 0$, namely $bid_i^* \leq wp_i$.
\end{itemize}

To sum up, for any $opportunity_i$, the bidding strategy would guarantee that the $x_i$ is optimal, which leads to the optimal solution of \eqref{lp1}. That is to say, when the optimal $x_i$ is 1 (the campaign should win $opportunity_i$), the bid price $bid_i$ based on the optimal bidding strategy is higher than $wp_i$, which would guarantee that the campaign wins $opportunity_i$. The reasoning is the same when the optimal $x_i$ is 0. Therefore, the campaign needs just to bid following the optimal bidding strategy in Eq. \eqref{eq_opt_func}, and the total advertising value would be maximized with the constraints.
\end{proof}

It is worth noting that Eq. \eqref{eq_opt_func} does not explicitly tell the value of $p$ and $q$. Actually, it is easy to derive the optimal $p$ and $q$ by solving the dual problem with well developed linear programming algorithms. Such work does not contribute to the comprehension of our work, so we do not discuss how to calculate the value of $p$ and $q$ here.  

\begin{figure}
    \centering
    \begin{subfigure}[b]{0.2\textwidth}
        \includegraphics[width=\textwidth]{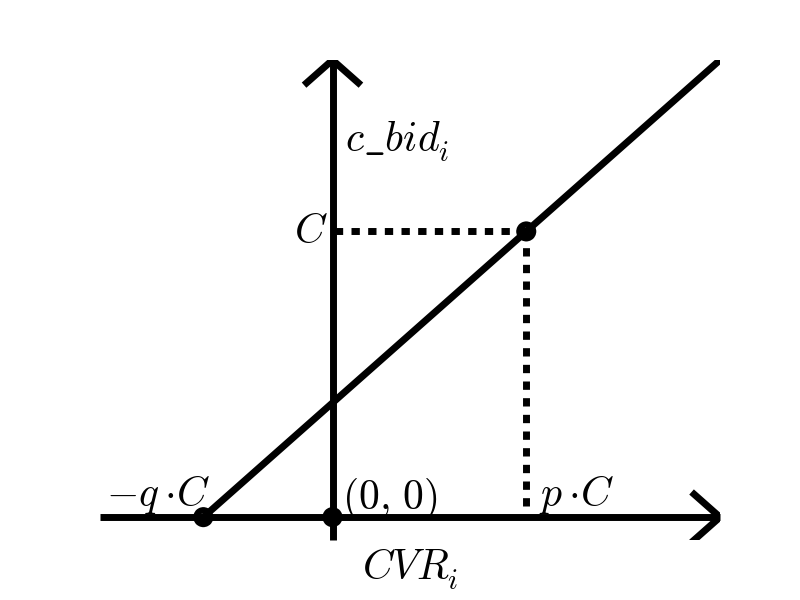}
        %\caption{Converging process}
        \caption{$c\_bid_i$}
        \label{cpc_op_function}

    \end{subfigure}
    \begin{subfigure}[b]{0.2\textwidth}
        \includegraphics[width=\textwidth]{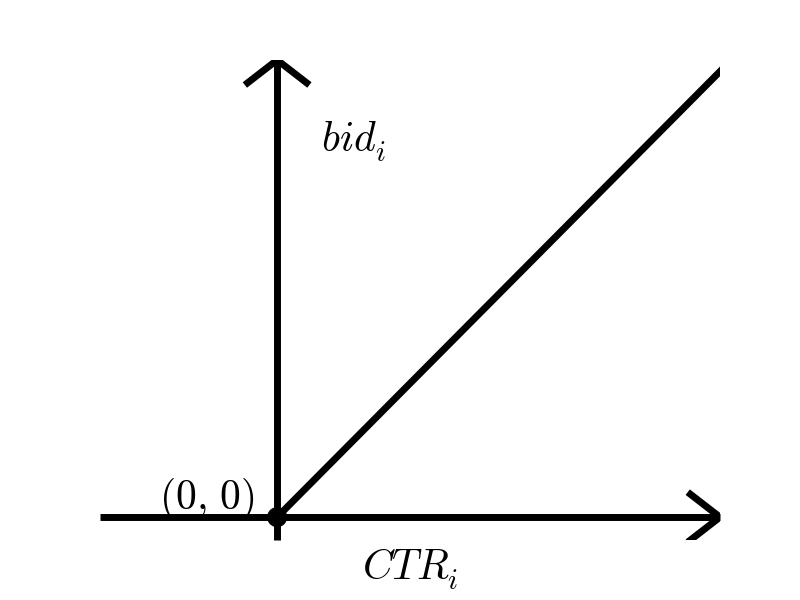}
        %\caption{Reward distribution}
        \caption{$bid_i$}
        \label{cpc_op_function_1}

    \end{subfigure}
    \caption{Optimal bidding strategy}			\label{fig:op_function}
\end{figure}

We reformulate the optimal bidding strategy for \eqref{lp1} into two stages as shown in Eq. \eqref{eq:cpc_bid}, Eq. \eqref{eq:cpc_bid_new} and Fig. \ref{fig:op_function}, where $c\_bid_i$ could be regarded as the bid price for a click. As an ad opportunity comes, we first determine the bid price for a click, i.e. $c\_bid_i$. After we determine $c\_bid_i$, the final bid price is calculated by multiplying $c\_bid_i$ with $CTR_i$, which could be automatically completed and is omitted in the following discussion. In addition, the cost for a click is naturally no higher than the bid price for a click under the GSP auction mechanism\cite{edelman2007internet}, so that $c\_bid_i$ is directly related to CPC. Therefore, we focus the discussion on the bid price for a click ($c\_bid_i$) instead of the final bid price to facilitate our demonstrations. 
\begin{align} 
c\_bid_i &= \frac{1}{p + q} \cdot CVR_i + \frac{q}{p + q} \cdot C & \label{eq:cpc_bid}\\
bid_i &= c\_bid_i \cdot CTR_i& \label{eq:cpc_bid_new}  \\
&= (\frac{1}{p + q} \cdot CVR_i + \frac{q}{p + q} \cdot C) \cdot CTR_i& \nonumber
\end{align}

We start the discussion on the optimal bidding strategy with some obvious facts as illustrated in Fig. \ref{cpc_op_function}. Firstly, the bid price for a click is strictly positive related to $CVR_i$. It does make sense that we should give a higher bid price for a more valuable click, and a lower price for a less valuable click. Secondly, the bid price is linear against $CVR_i$. It is a natural result since we are maximizing the sum of the value, which is a linear function against $CVR_i$ itself. Thirdly, the bidding strategy would definitely cross two points as emphasized in the figure. We shall discuss it in details in later analysis. However, there is an unusual fact: unlike the widely adopted bidding strategies \cite{zhang2016feedback, perlich2012bid}, the function does not necessarily get through the origin. Specifically, the bid price would be non-zero even though the ad opportunity has no value for the advertiser. It is a little bit unintuitive that we bid a non-zero price for an ad opportunity without value. We state the reason as follows: considering a CPC constraint is given, the bidding strategy tries to win some cheap ad opportunity, even without any value, to lower the overall CPC, and thus to win some valuable ad opportunity with high CPC. 

\section{System Design}
		
	In this section, we address the applicability issue of the bidding strategy in the industrial scenario, and propose the feedback control-based solution. Based on the analysis of the hyper parameters in the optimal bidding strategy, we present the multivariable control system. 
	
\subsection{Applicability Issue}
	As shown in \eqref{lp2}, in order to solve the linear programming problem and obtain the optimal bidding strategy, we need to know exactly the information of every ad opportunity of the day, including the winning price, CTR and CVR. Such information, however, could not be obtained in real world until the end of the day, while the optimal bidding strategy need to be determined before the campaign starts. Apart from the fact that it is difficult to estimate all of them in the impression level precisely before the campaign starts\cite{wang2016functional, zhou2018deep, yang2017bayesian}, how to predict the ad opportunity itself of the day is still an open question due to the dynamic environment\cite{snoddy2018system}. One may argue that there are many statistical algorithms proposed to solve such a problem: the optimal solution could be derived from sufficient historical data and be applied in the future. One strong assumption made in such algorithms is that the distribution of the variables are stationary. However, the distribution of not only the ad opportunities, but also other factors such as the winning price, CTR and CVR are not stationary in the dynamic RTB environment. Therefore, the optimal bidding strategy derived on historical data becomes no longer optimal for future, and may even break the CPC constraint. 
%To show the dynamic environment in RTB, we illustrate the the impression volume across days and the market price across hours for an advertiser in Fig. \ref{fg:volatility}, where the data is from Taobao.com. Given the instability of the bidding environment, propose the feedback control-based solution.
%
%\begin{figure}
%    \centering
%    \begin{subfigure}[b]{0.2\textwidth}
%        \includegraphics[width=\textwidth]{rtb_pv.png}
%        %\caption{Converging process}
%        \caption{Impression volume}
%
%    \end{subfigure}
%    \begin{subfigure}[b]{0.2\textwidth}
%        \includegraphics[width=\textwidth]{rtb_cpm.png}
%        %\caption{Reward distribution}
%        \caption{Market price}
%
%    \end{subfigure}
%    \caption{The instability of the bidding environment}\label{fg:volatility}
%\end{figure}

\subsection{Feedback Control-based Solution} 

	As discussed in the last section, due to the dynamic bidding environment, the bidding strategy we derive on the historical data may be unreliable. Therefore, we need to leverage the real-time information to adjust the bidding strategy. Feedback control, which deals with the dynamic systems from feedback and outside noise\cite{kumar2014control}, is widely adopted in the industry because of its robustness and effectiveness. A feedback control system is to achieve desirable performance by adjusting the system input based on the feedback of the system output. In our scenario, we could naturally integrate the bidding strategy and the RTB environment as a dynamic system, and regard the hyper parameters of the bidding strategy $p$ and $q$ as the input of the system. By doing so, the problem is transformed into a feedback control problem. There is still one problem: what is the desirable performance, and consequently what feedback of the output should we care about? Firstly, we aim to maximize the advertising value and control CPC. Secondly, in order to maximize the advertising value, we should pace the budget spending to win the valuable ad opportunities scattered across time\cite{song2017volume}. Therefore, we need to pace the budget spending, and control CPC simultaneously. We propose the feedback solution as follows: to improve the advertising performance, we pace the budget spending and control CPC based on their real-time feedback.

\begin{figure}
 	\centering    
    \includegraphics[width=0.48\textwidth]{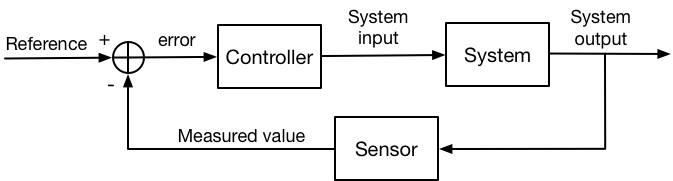}
    \caption{A block diagram of the feedback control system}
    \label{fg:feedback_control}
\end{figure}
		
We briefly introduce the standard feedback control system. A block diagram of the feedback control system is illustrated as Fig. \ref{fg:feedback_control}. The desired value of the ouput is called the reference, which is pre-set depending on the specific task. The sensor measures the actual value of the variable from the system output, and transmit it to the controller. Comparing the measured value and the reference, the controller would adjust the input of the system by its pre-defined algorithms or strategies to diminish the difference between them.  Proportional-Integral-Derivative (PID) controller\cite{bennett1993development} is the most widely adopted feedback controller in the industry. It is known that a PID controller delivers best performance in the absence of knowledge of the underlying process \cite{benner1993history}.  A PID controller continuously calculates the error $e(t)$ between the measured value $y(t)$ and the reference $r(t)$ at every time step $t$, and produce the control signal $u(t)$ based on the combination of proportional, integral, and derivative terms of $e(t)$. The control signal $u(t)$ is then sent to adjust the system input $x(t)$ by the actuator model $\phi(x(0), u(t))$. It is practical and common to use discrete time step ($t_1, t_2, ...$) in online advertising scenario, so the process of PID could be formulated as following equations, where $k_p$, $k_i$, and $k_d$ are the weight parameters of a PID controller. 
	
\begin{equation} \label{pid_et}
\begin{aligned}
e(t) = r(t) - y(t) \ \ \\
\end{aligned}
\end{equation}	
	
\begin{equation} \label{pid_ut}
\begin{aligned}
u(t) = k_p  e(t) + k_i  \sum\limits_{i=1...t}e(k) + k_d  (e(t) - e(t-1)) \ \ \\
\end{aligned}
\end{equation}

\begin{equation} \label{pid_xt}
\begin{aligned}
x(t+1) = \phi (x(0), u(t)) \ \ \\
\end{aligned}
\end{equation}

\subsection{Analysis on Hyper Parameters}

We have transformed the problem into a feedback control problem, and determined the dynamic system (the bidding strategy and RTB), the input parameters ($p$ and $q$), and the output variables (budget spending and CPC) in last sections. The challenge is how to control the budget spending and CPC simultaneously by adjusting $p$ and $q$. Due to the multiple input parameters and the multiple output variables, we cannot directly apply the PID controllers in our scenario since such controllers are designed for the system with a single input parameter and a single output variable. There has been some multivariable control methods such as model predictive control \cite{rawlings2009model} proposed to deal with such a system, and we shall leverage their underlying idea in our design in the next section. In this section, we revisit the optimal bidding strategy in Eq. \eqref{eq:cpc_bid} and share our ideas to design the multivariable control system.

We analyse how the hyper parameters $p$ and $q$ contribute to the bidding strategy in Eq. \eqref{eq:cpc_bid}. Please recall that $p$ and $q$ are dual variables incorporated by the constraint \eqref{con_p} and \eqref{con_q} respectively, and we explore their relationship with the corresponding constraint.

\begin{figure}
    \centering
    \begin{subfigure}[b]{0.2\textwidth}
        \includegraphics[width=\textwidth]{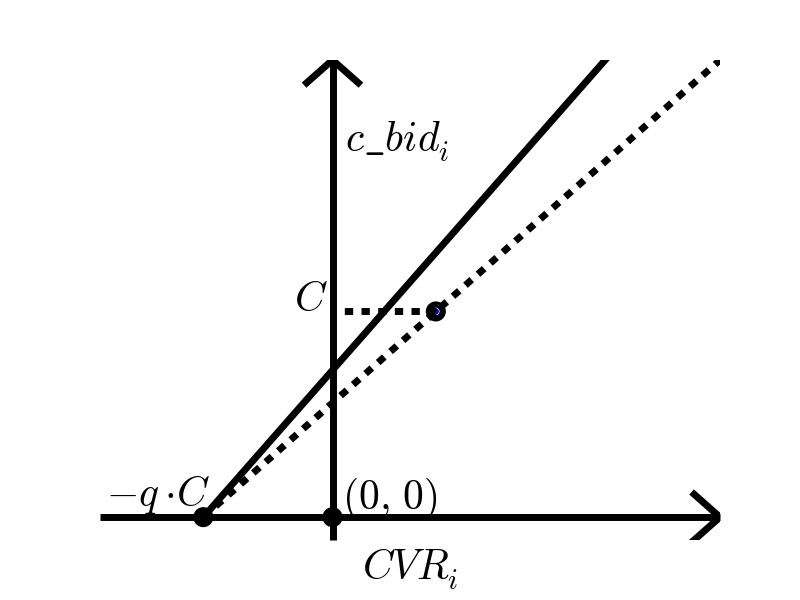}
        %\caption{Converging process}
        \caption{$p$ decreased}

        \label{fig:reward_net_converge_cmp}
    \end{subfigure}
    \begin{subfigure}[b]{0.2\textwidth}
        \includegraphics[width=\textwidth]{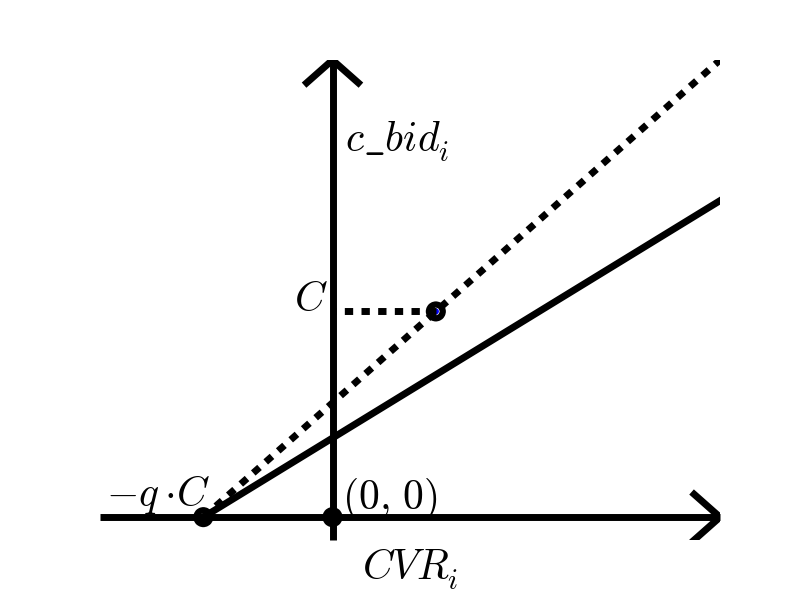}
        %\caption{Reward distribution}
        \caption{$p$ increased}

        \label{fig:reward_dist}
    \end{subfigure}
    \begin{subfigure}[b]{0.2\textwidth}
        \includegraphics[width=\textwidth]{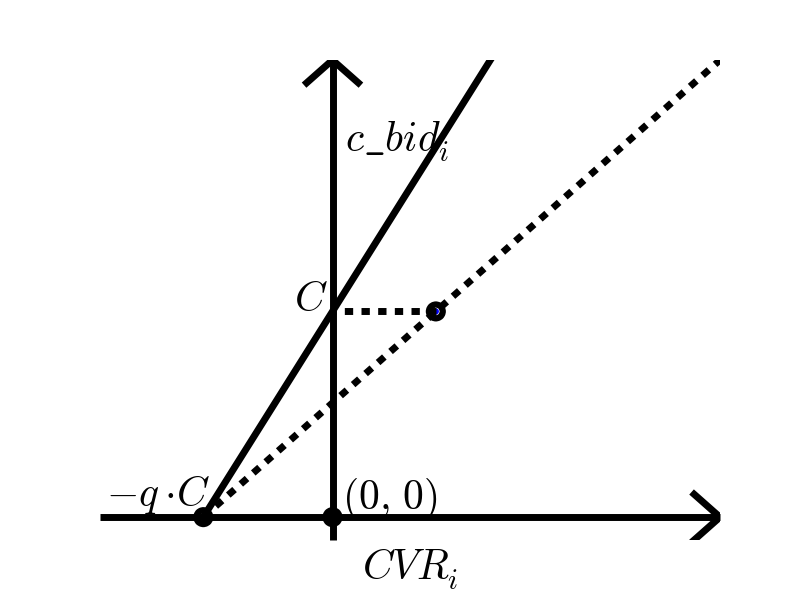}
        %\caption{Reward distribution}
        \caption{$p = 0$}

        \label{fig:reward_dist}
    \end{subfigure}
    \begin{subfigure}[b]{0.2\textwidth}
        \includegraphics[width=\textwidth]{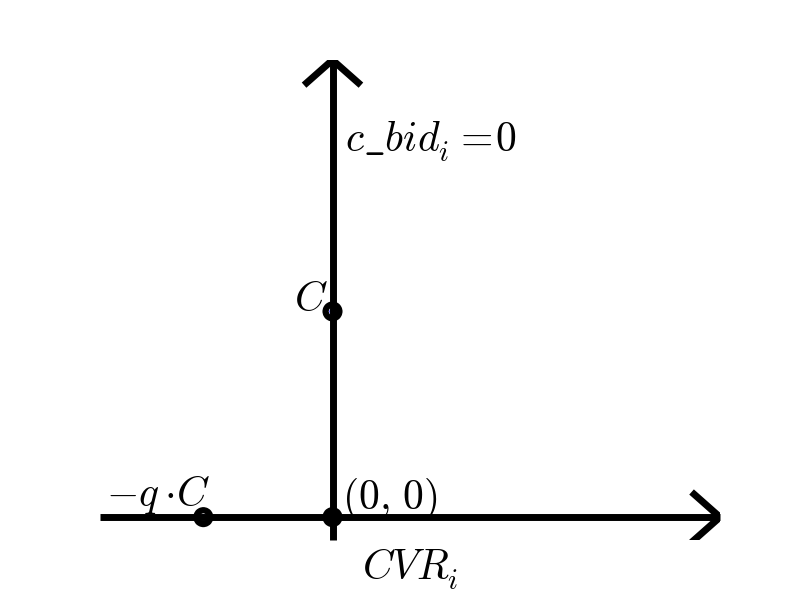}
        %\caption{Reward distribution}
        \caption{$p = \infty$}

        \label{fig:reward_dist}
    \end{subfigure}
    \caption{Bidding strategies with $q$ fixed and $p$ respectively decreased, increased, equal to $0$, and equal to $\infty$. The dashed line is illustrated as a reference to the original function.}\label{fg:p_variate}
\end{figure}

Fig. \ref{fg:p_variate} shows the optimal bidding strategy with $q$ fixed and $p$ respectively decreased, increased, equal to $0$ and equal to $\infty$. It is worth noting that the bidding price would exactly influence the expected cost: the higher bidding price leads to more cost since the campaign may win more ad opportunities. As illustrated in Fig. \ref{fg:p_variate}, the bidding price would be generally lowered or lifted as $p$ increases or decreases. When we increase $p$, the bidding strategy would rotate clockwise around the point $(-q \cdot C,\ 0)$. The resulting fact is that: the bidding price that is above zero would be lowered, and the bidding price that is below zero\footnote{Although the bidding price would never be a negative value, the analytic continuation would help to comprehend the mathematics property.} would be lifted, and thus the expected cost would be lessened. Please take $p = \infty$ and $p = 0$ as the extreme examples. The advertising campaign would never be charged when its bid price constantly equals to zero with $p = \infty$. When $p = 0$, the budget becomes no longer a constraint. The bidding price does not become infinitely high because there is still a CPC constraint and the slope of the bidding strategy is controlled completely by $q$. If we simultaneously set $q$ to be 0, which means the CPC constraint also is removed, the bidding price would be infinitely high and the campaign would win all ad opportunities. According to the illustrations, we claim that $p$ has a direct and effective control capability on the budget spending, and could definitely lower the budget spending speed.

\begin{figure}
    \centering
    \begin{subfigure}[b]{0.2\textwidth}
        \includegraphics[width=\textwidth]{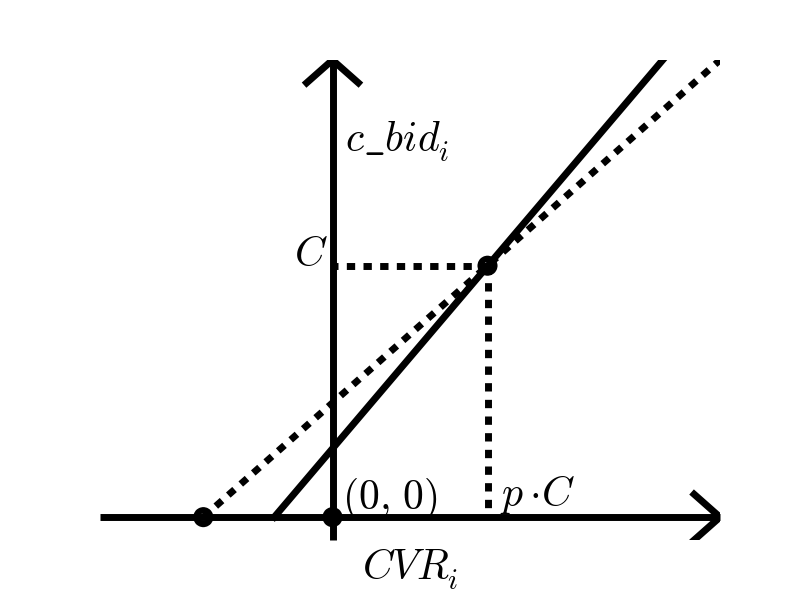}
        %\caption{Converging process}
        \caption{$q$ decreased}

        \label{fig:reward_net_converge_cmp}
    \end{subfigure}
    \begin{subfigure}[b]{0.2\textwidth}
        \includegraphics[width=\textwidth]{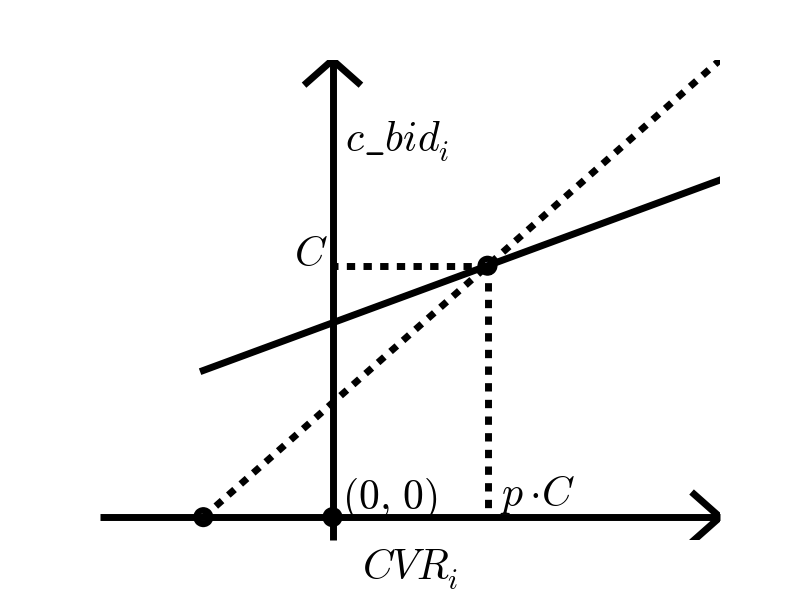}
        %\caption{Reward distribution}
        \caption{$q$ increased}

        \label{fig:reward_dist}
    \end{subfigure}
    \begin{subfigure}[b]{0.2\textwidth}
        \includegraphics[width=\textwidth]{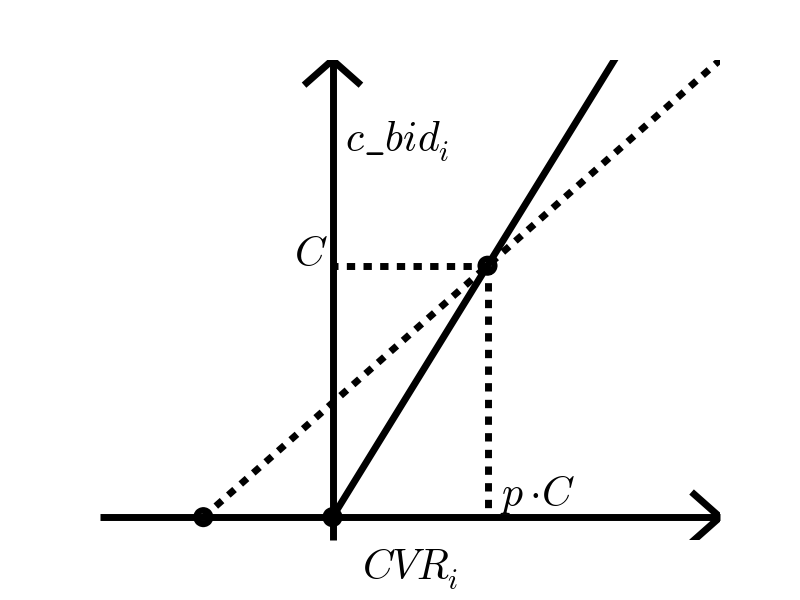}
        %\caption{Reward distribution}
        \caption{$q = 0$ }

        \label{fig:reward_dist}
    \end{subfigure}
    \begin{subfigure}[b]{0.2\textwidth}
        \includegraphics[width=\textwidth]{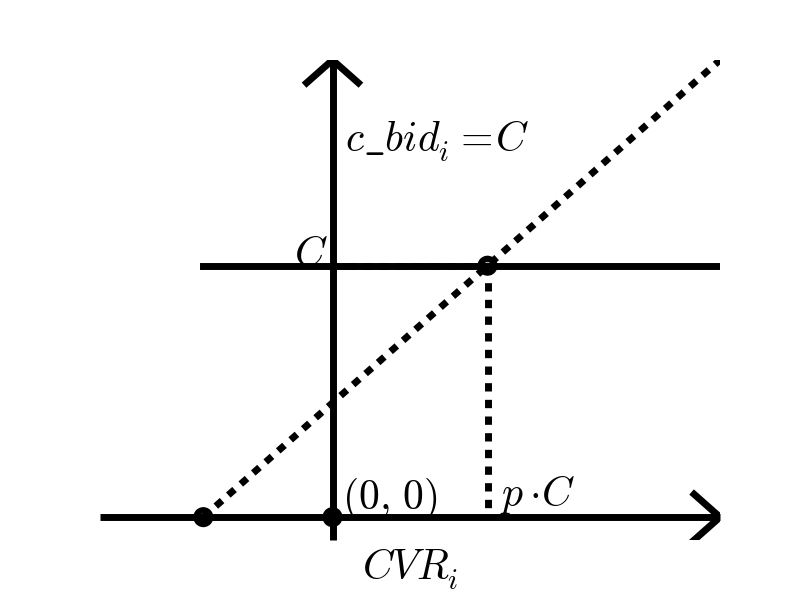}
        %\caption{Reward distribution}
        \caption{$q = \infty$}

        \label{fig:reward_dist}
    \end{subfigure}
    \caption{Bidding strategies with $p$ fixed and $q$ respectively decreased, increased, equal to $0$ and equal to $\infty$. The dashed line is illustrated as a reference to the original function.}\label{fg:q_variate}
\end{figure} 

In a similar way, we fix $p$ and set $q$ to be respectively decreased, increased, equal to 0, and equal to $\infty$. As illustrated in Fig. \ref{fg:q_variate}, $q$'s effects on the optimal bidding strategy is notably different to that of $p$. When we increase or decrease $q$, the bidding strategy would rotate clockwise or counter-clockwise around the the point $(p \cdot C,\ C)$. Taking increasing $q$ for example, the bid price above $C$ would be lowered, and that below $C$ would be lifted. Thus the campaign would win more ad opportunities whose CPC is below $C$, and less ad opportunities whose CPC is above $C$. The composite result is that the CPC is more likely to be below $C$. In the extreme cases, CPC would be guaranteed below $C$ when $q = \infty$. When $q$ is set to be zero, which means the CPC constraint is removed, the bidding strategy is determined by $p$ and degenerates to the optimal budget-constrained bidding strategy in \cite{wu2018budget, zhang2016feedback, perlich2012bid}. Based on the analysis, we claim that CPC could be definitely controlled by $q$. We propose the following two statements: 
\begin{enumerate}[align=right,leftmargin=0.17in]
\item Hyper parameter $p$ has a direct and effective control capability on the budget spending, and the budget spending speed could be definitely lowered by $p$ given whatever value of $q$.
\item Hyper parameter $q$ has a direct and effective control capability on CPC, and the CPC constraint could be definitely achieved by adjusting $q$ given whatever value of $p$.
\end{enumerate}

\subsection{Multivariable Control}

\begin{figure}
 	\centering    
    \includegraphics[width=0.44\textwidth]{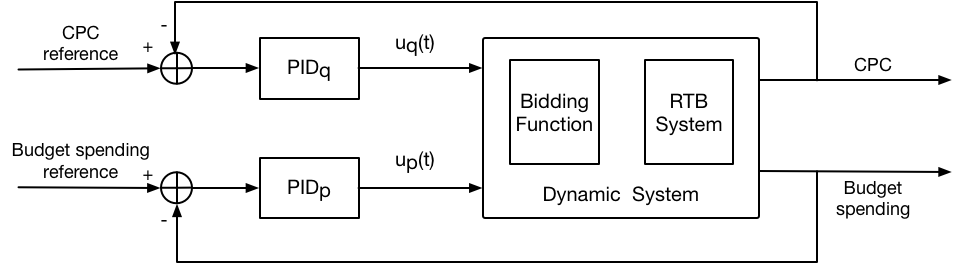}
    \caption{Independent PID control system}
    \label{fg:independent_pid}
\end{figure}

As we state in the last section, the budget spending and CPC could be definitely controlled by $p$ and $q$ respectively. In other words, $p$ and $q$ could be used to control the budget spending and CPC independently, and regard each other as outside noise. Therefore, we could decompose the multivariable feedback control problem into two single-variable feedback control problem. By doing so, PID controllers could be easily deployed, and we propose the independent PID design in Fig. \ref{fg:independent_pid}, where and afterwards we slightly abuse the subscript $p$ and $q$ to differentiate the two controllers. 

\begin{figure}
 	\centering    
    \includegraphics[width=0.48\textwidth]{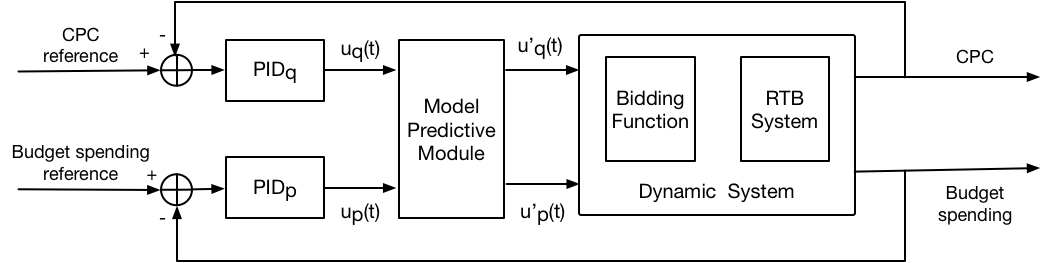}
    \caption{Model predictive PID control system}
    \label{fg:mpc_pid}
\end{figure}

To make a further step, we revisit the analysis in Section 3.3. As illustrated in Fig. \ref{fg:p_variate}, increasing $p$ would generally lower the bidding price and reduce the budget spending speed. Another non-trivial effect caused by increasing $p$ is that the expected CPC would also be lowered. According to such observations, adjusting $p$ actually casts influence on CPC. Similarly, adjusting $q$ also casts influence on budget spending. Although our independent PID control system could deal with such coupling effect, which is treated as outside noise by controllers, it is possible to further improve the performance of the system by addressing such an issue. However, such coupling effect is hard to be quantified and compensated since we have no explicit knowledge of the dynamic system. In order to address such a problem, we leverage the underlying idea of model predictive control (MPC)\cite{rawlings2009model} to predict and compensate the coupling effect. It is worth mentioning that we do not directly apply MPC in our control system since modelling the highly non-linear RTB environment is costly and even impractical. In our design, combined with the human knowledge, the model predictive module needs only to predict the coupling effect, which could be approximated by a linear model. As illustrated in Fig. \ref{fg:mpc_pid}, a model predictive module is deployed after PID controllers to regulate the control signal by addressing the coupling effect. 

One of the most important components of MPC is a model to represent the behaviour of the dynamic system. In our case, we model the bidding environment with respect to the cost and CPC as shown in Eq. \eqref{env_model_matrix}, where $\textbf{X}$ is a $2 \times 2$ matrix and $\textbf{b}$ is a $2 \times 1$ matrix. After we obtain the expected $\Delta cost$ and $\Delta CPC$ from the feedback, we could adjust $p$ and $q$ by solving the equation in Eq. \eqref{env_model_delta_set}, and derive the result as shown in Eq. \eqref{env_model_delta_res}. The formulation in Eq. \eqref{env_model_delta_res} demonstrates that the control signal of $p$ and $q$ should be a linear combination of the changes of cost and CPC. Therefore, we define the model predictive module by Eq. \eqref{env_model_mpm}, where $\Delta cost$ and $\Delta CPC$ are quantified by $u_p(t)$ and $u_q(t)$ respectively, and $[\textbf{X}]^{-1}$ is approximated by a $2 \times 2$ matrix determined by $\alpha$ and $\beta$. By approximating $[\textbf{X}]^{-1}$, we could simply regard $\alpha$ and $\beta$ as two weight parameters and search their best value in the training set. Although such an approximation undermines the capability to represent the system, it makes the controller more robust and stable against the changing environment. It is worth mentioning that we propose the matrix $\textbf{X}$ and $\textbf{b}$ to model the dynamic system, however, the exactly value of such matrices is not explicitly required. As shown in Eq. \eqref{env_model_mpm}, we take advantage of such matrices to address the coupling effect, and obtain the approximated function determined by only $\alpha$ and $\beta$.

\begin{equation} \label{env_model_matrix}
\left[\begin{array}{cccc} 
    cost \\
    CPC 
\end{array}\right] = 
\left[\begin{array}{cccc} 
    \textbf{X}\;\; \textbf{b} 
\end{array}\right]
\left[\begin{array}{cccc} 
    p \\
    q \\
    1
\end{array}\right]
\end{equation}
\begin{equation} \label{env_model_delta_set}
\left[\begin{array}{cccc} 
    \Delta cost \\
    \Delta CPC 
\end{array}\right] = 
\left[\begin{array}{cccc} 
    \textbf{X}
\end{array}\right]
\left[\begin{array}{cccc} 
    \Delta p \\
    \Delta q 
\end{array}\right]
\end{equation}
\begin{equation} \label{env_model_delta_res}
\left[\begin{array}{cccc} 
    \Delta p \\
    \Delta q 
\end{array}\right] = 
{\left[\begin{array}{cccc} 
    \textbf{X}
\end{array}\right]}^{-1}
\left[\begin{array}{cccc} 
    \Delta cost \\
    \Delta CPC 
\end{array}\right]
\end{equation}
\begin{equation} \label{env_model_mpm}
\left[\begin{array}{cccc} 
    u'_p(t) \\
    u'_q(t) 
\end{array}\right] = 
\left[\begin{array}{cccc} 
    \alpha \;\;\; 1-\alpha \\
    1-\beta \;\;\; \beta
\end{array}\right]
\left[\begin{array}{cccc} 
    u_p(t) \\
    u_q(t) 
\end{array}\right]
\end{equation}

\section{Empirical Study}
		
	In this section, we conduct comprehensive experiments to demonstrate our statements and the advantage of our multivariable control systems. After describing the real dataset and related metrics, we present the implementation details. Experiments are conducted on the example campaigns to prove the the claimed control capability of the hyper parameters. To show the superiority of our system, we compare our methods with the state of the art in the industry practices on a large group of campaigns. 
	
\subsection{Experiment Setup}

\subsubsection{Dataset}
We base the experiments on the real dataset from Taobao.com. The dataset consists of 40 advertising campaigns across continuous days and totally $20M$ bid logs. It is split according to the date as training dataset and test dataset. The key information of the dataset from the perspective of a specific advertising campaign could be summarized as Tab. \ref{tab:dataset}. We mainly use the information of the winning price, $CTR$ and $CVR$. The winning price of each ad opportunity corresponding to the specific campaign is recorded after each online auction ends. Since Taobao.com is also a publisher, the winning price could be observed even the advertiser missed the ad opportunity in the online auction. $CTR$ and $CVR$ are estimated by the online deployed models, which leverage extensive realtime and historical information of the user and ads. Please refer to \cite{zhou2018deep} for details of the online deployed estimation models.
\begin{table}[h]
\centering
\caption{Key information of the dataset}\label{tab:dataset}
\begin{tabular}{cc}
\hline
Key & Description \\ \hline
$opportunity_i$ 	& identity of the ad opportunity\\ %\hline
$timestamp_i$ 	& arriving time of $opportunity_i$\\ %\hline
$wp_i$			& winning price of $opportunity_i$ \\ %\hline
$CTR_{i}$			& estimated click-through rate of $opportunity_i$ \\ %\hline
$CVR_{i}$			& estimated conversion rate of $opportunity_i$	\\ \hline
\end{tabular}
\end{table}

\subsubsection{Metrics}

The goal of the bidding strategy and system is to maximize the total value of the winning ad opportunities, and control CPC under the given threshold. We quantify the advertising value by the sum of $CTR \cdot CVR$, which corresponds to the expected outcomes of the conversion. It is worth mentioning that we regard $CTR \cdot CVR$ as the value, instead of the actual conversion event, to exclude the inaccuracy caused by the estimation models. Even though some previous work evaluates the bidding strategy by the actual conversions, we argue that the estimation error actually cast non-trivial influence on the results. A campaign with a fixed bidding strategy may gain more clicks/conversions just by optimizing the estimation models, so we regard $CTR \cdot CVR$ as the true conversion to diminish such influence. 
\begin{enumerate}[align=right,leftmargin=0.17in]
\item $R$ represents the advertising value of the campaign.
\item $R^*$ represents the maximum advertising value the campaign could achieve with the budget and CPC constraints.
\item $R/R^*$ could be used to evaluate how close the advertising performance is to the ideal result. 
\item $CPC_{ratio}$ is the proportion of the campaigns that satisfy the CPC constraint (overshoot within $10\%$ is allowed), which could be used to evaluate the CPC control capability when comparing different methods with each other on a large group of campaigns. 
\item $Value_{ratio}$ is the average $R/R^*$ on the campaigns whose CPC constraint holds, which is to evaluate the advertising value achievement. As for those campaigns that break the CPC constraint, we exclude their $R/R^*$ when we calculate $Value_{ratio}$ since wining more value by breaking the CPC constraint is not allowed in our scenario.
\end{enumerate}

\subsubsection{Implementation Details}
We adopt the actuator shown in Eq. \eqref{pid_xt_sst} in the PID controllers and the baseline strategies, where the sign of $u(t)$ depends on the relationship between the input parameter and the output variable. In addition, it needs to be noted that we actually care about the accumulated CPC when the campaign ends, and the real-time CPC of each time step contributes differently to the accumulated CPC because the quantity of clicks in each time step is different. The traditional PID error could not address the different weight in our scenario, so we weigh the error by the quantity of clicks and modify the control signal for $q$ as shown in Eq. \eqref{pid_ut_q_error} and Eq. \eqref{pid_ut_q}, where $u(t)$ is calculated by Eq. \eqref{pid_ut} and $click(t)$ represents the quantity of clicks in the time step $t$. By such a modification, the PID controller would constantly increase its attention on the accumulated CPC, and give each time step different weights.

%the CPC of each time step contributes differently to the accumulated CPC because the quantity of clicks in each time step is different. Therefore, we give different weight to the control signal of $q$ according to the quantity of clicks as shown in Eq. \eqref{pid_ut_q}, where $u(t)$ is calculated by Eq. \eqref{pid_ut} and $click(t)$ represents the quantity of clicks in the time step $t$. Similarly, we adjust $p$ in the same way as shown in Eq. \eqref{pid_ut_p}, where $cost(t)$ represents the cost in time step $t$. By such a modification, we could effectively reduce the influence caused by the time steps with little significance and thus enhance the controller's performance.

To determine the weight parameters of a PID controller, as well as the weight parameters $\alpha$ and $\beta$, we grid-search the best setting on the training dataset, and apply it on the test dataset. We regard every one hour as a time step, so the maximum time step $T$ equals to $24$. 
\begin{equation} \label{pid_xt_sst}
\begin{aligned}
x(t+1) = x(0) \cdot exp(-u(t))\ \ \\
\end{aligned}
\end{equation}
\begin{equation} \label{pid_ut_q_error}
\begin{aligned}
e_q(t) = click(t) \cdot (r(t) - y(t))  \ \ \\
\end{aligned}
\end{equation}
\begin{equation} \label{pid_ut_q}
\begin{aligned}
u_q(t) = \frac{1}{\displaystyle\sum\limits_{i=1...t} click(t)} \cdot u(t)  \ \ \\
\end{aligned}
\end{equation}
%\begin{equation} \label{pid_ut_p}
%\begin{aligned}
%u_p(t) = \frac{t \cdot cost(t)}{\displaystyle\sum\limits_{i=1...t} cost(t)} \cdot u(t)  \ \ \\
%\end{aligned}
%\end{equation}

As we discussed in last sections, the PID controllers need a reference. In our experiments, CPC given by the advertiser is set as the constant reference to control $q$. Considering the ad opportunity and winning price, which have an immediate impact on the cost, show different statistical characteristics across time steps,  the cost reference should be customized with respect to the time step. We calculate the cost distribution on the training dataset, which is the ideal cost of each time step normalized by the total cost of the day, as the cost reference. Our experiment flow steps are as follows: 1) optimal $p$ and $q$ are calculated on the training dataset; 2) the bidding process is simulated on the test dataset, where the calculated $p$ and $q$ are applied as the initial hyper parameters; 3) the simulation would finish when the campaign runs out of budget or there is no more ad opportunity.

\subsection{Control Capability}
\begin{figure}
    \centering
    \begin{subfigure}[b]{0.2\textwidth}
        \includegraphics[width=\textwidth]{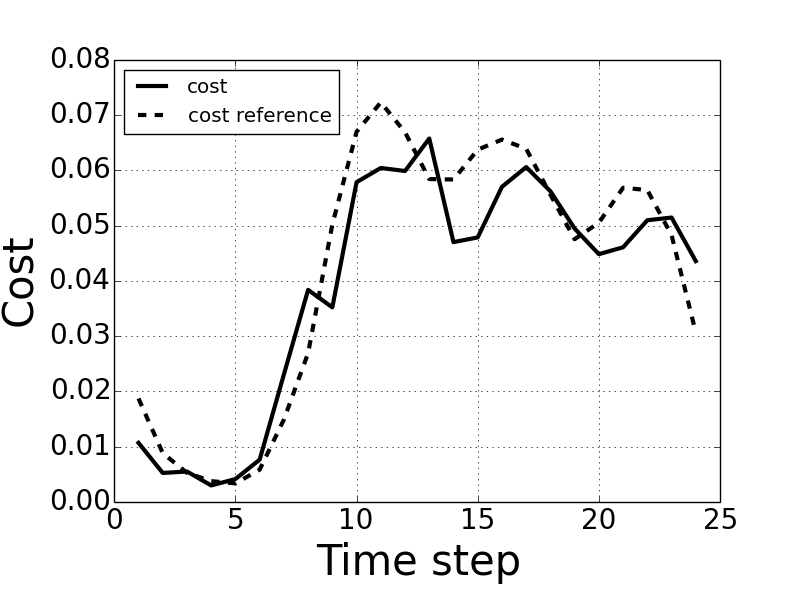}
        %\caption{Converging process}
        \caption{Cost per time step}

        \label{fig:cost_per_step}
    \end{subfigure}  
    \begin{subfigure}[b]{0.2\textwidth}
        \includegraphics[width=\textwidth]{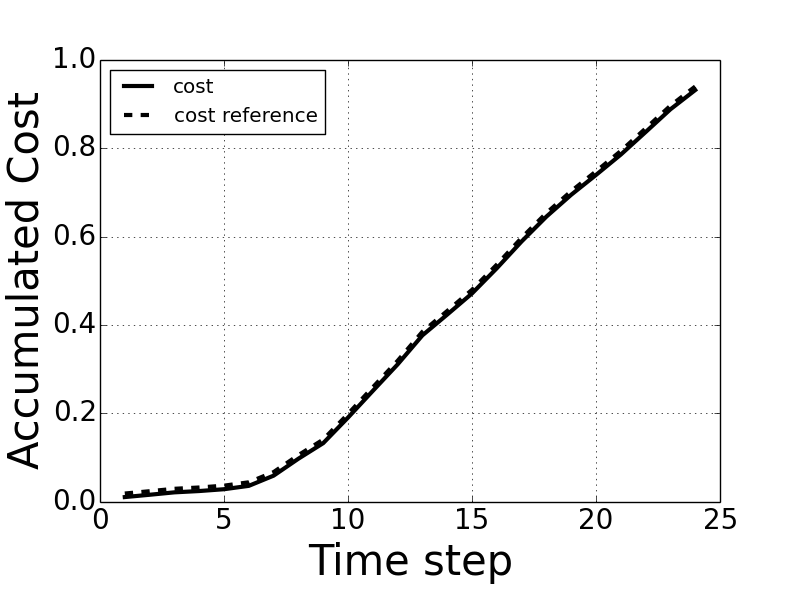}
        %\caption{Reward distribution}
        \caption{Budget spending}

        \label{fig:accumulated_cost}
    \end{subfigure}
    
%    \begin{subfigure}[b]{0.2\textwidth}
%        \includegraphics[width=\textwidth]{control_capability_p_1.png}
%        %\caption{Reward distribution}
%        \caption{Correlation of $p$ and cost}
%        \label{fig:correlation_p_cost}
%	\end{subfigure}

    \caption{Control performance on budget spending. The cost distribution is set as the cost reference.}\label{fig:control_cap_p}
\end{figure}

\begin{figure}
    \centering
    \begin{subfigure}[b]{0.2\textwidth}
        \includegraphics[width=\textwidth]{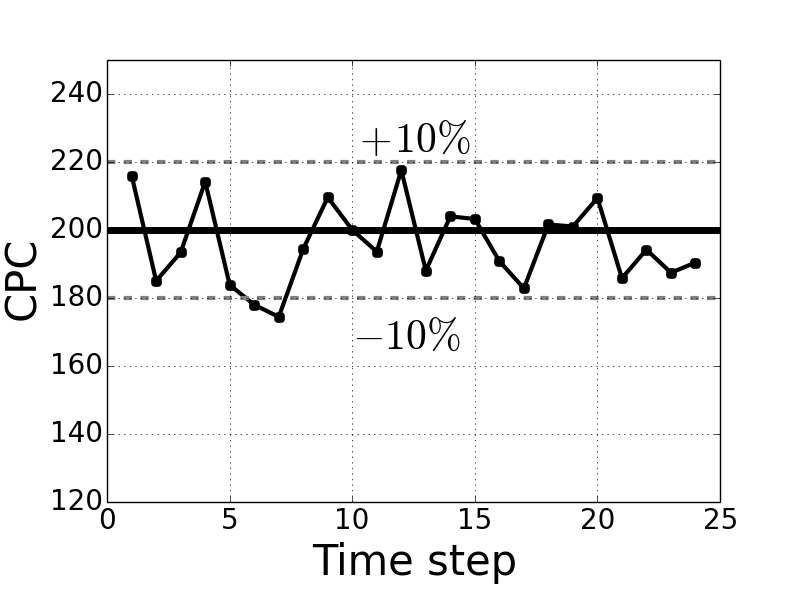}
        %\caption{Converging process}
        \caption{CPC per time step}

        \label{fig:cpc_per_step}
    \end{subfigure}  
    \begin{subfigure}[b]{0.2\textwidth}
        \includegraphics[width=\textwidth]{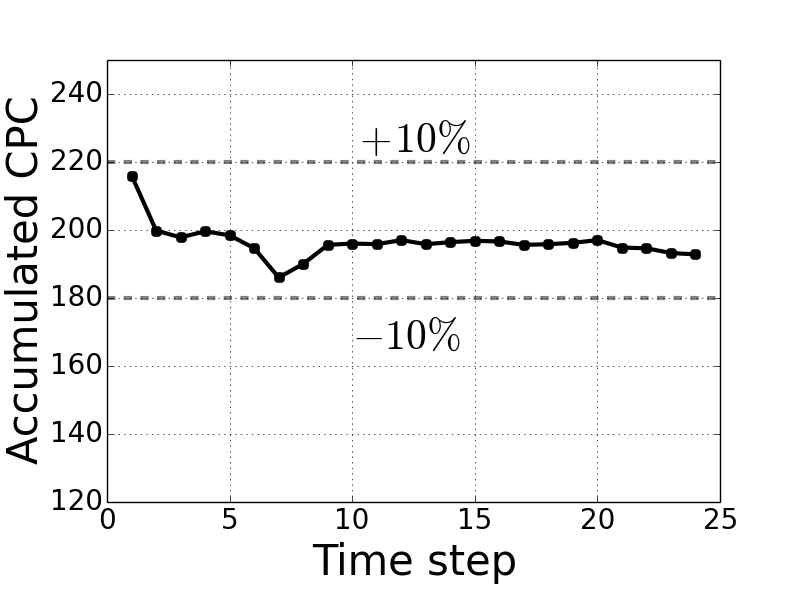}
        %\caption{Reward distribution}
        \caption{Accumulated CPC}

        \label{fig:accumulated_cpc}
    \end{subfigure}
    
%    \begin{subfigure}[b]{0.2\textwidth}
%        \includegraphics[width=\textwidth]{control_capability_q_1.png}
%        %\caption{Reward distribution}
%        \caption{Correlation of $q$ and CPC}
%        \label{fig:correlation_q_cpc}
%	\end{subfigure}

    \caption{Control performance on CPC. The CPC reference is set to be 200, and the $\pm10\%$ interval around the reference value is defined as the tolerable margin.}\label{fig:control_cap_q_cpc}
\end{figure}

In this section, we do experiments to demonstrate our statements that budget spending and CPC could be independently controlled by $p$ and $q$. We adjust both hyper parameters simultaneously without the model predictive module and illustrate the control performance in terms of budget spending and CPC respectively on the example campaign. The controlling performance is illustrated in Fig. \ref{fig:control_cap_p} and Fig. \ref{fig:control_cap_q_cpc}

As illustrated in Fig. \ref{fig:control_cap_p}, the cost of each time step controlled by $p$ fluctuates around the cost reference, and the accumulated cost is well controlled referring to the accumulated cost reference. As Fig. \ref{fig:control_cap_q_cpc} shows, CPC of each time step gets quickly confined within the tolerable margin, and the accumulated CPC is successfully controlled under the given reference. It is worth mentioning that the real-time CPC shows observable fluctuation across time steps since the attention on the real-time CPC is gradually decreased in our design. Compared with the real-time CPC, the accumulated CPC delivers stable performance as illustrated in Fig. \ref{fig:accumulated_cpc}. According to the experiments, although $p$ and $q$ have interference with each other, they could be independently adjusted to control the corresponding output variables. 

%CPC and $q$ show strong correlation according to Fig. \ref{fig:correlation_q_cpc}: their value tends to change in the opposite directions. Such results verify our statements that $q$ has a direct and effective control capability on CPC.

\subsection{Performance Evaluation}

In this section, we compare our approach with the state of the art in the industry. We first introduce our baseline strategies, and evaluate all of them on the real-world dataset. 

\subsubsection{Baseline Strategies}

\begin{enumerate}[align=right,leftmargin=0.17in]
\item \textbf{Cost-min}: Cost-min\cite{kitts2017ad} is a generic algorithm to address multiple constraints in advertising scenario, and could be applied in our scenario as follows. The bidding strategy Eq. \eqref{eq:budget_bid} is adopted. We adjust $b_0$ by a PID controller according to the cost reference, and set the upper bound of $c\_bid_i$ to be $C$. The CPC constraint of the advertising campaign would always hold with Cost-min because of the truncated bid price. We divide the given CPC by the averaged CVR on all bid logs of the campaign to initialize $b_0$.
\item \textbf{Fb-Control}: Zhang et al. proposed a feedback control mechanism that dynamically adjusts the bids to control CPC\cite{zhang2016feedback}. In their work, they adopted the generalised bidding strategy, which is shown in Eq. \eqref{eq:budget_bid_raw}, and constantly adjust $b_0$ by a PID controller according to the feedback of CPC. The given CPC is set as the initial value of $b_0$. For simplicity, we reference their method as Fb-Control.
\item \textbf{Fb-Control-M}: Fb-Control does not consider the value of a click (i.e. $CVR$), which is important to improve the advertising performance. Therefore, we modify their bidding strategy to better fit into our scenario as shown in Eq. \eqref{eq:budget_bid}, where $b_0$ is adjusted by a PID controller according to our cost reference, and the upper bound of $c\_bid_i$ is controlled by an independent PID controller according to the feedback of CPC. The upper bound, initialized by the given CPC, would truncate $c\_bid_i$ every time $c\_bid_i$ exceeds the value, and $b_0$ is initialized in the same way with Cost-min. We reference the modified method as Fb-Control-M. 
\end{enumerate}  
\begin{equation} \label{eq:budget_bid}
\begin{aligned}
c\_bid_i = b_0 \cdot CVR_i \\
\end{aligned}
\end{equation}	
\begin{equation} \label{eq:budget_bid_raw}
\begin{aligned}
c\_bid_i = b_0 \\
\end{aligned}
\end{equation}	
\subsubsection{Experimental Results}

We reference our independent PID control system as \textbf{I-PID}, and the model predictive PID control system as \textbf{M-PID} for simplicity. We compare I-PID and M-PID with strategies introduced in the last section on the real-word dataset, and the result on 40 test campaigns is shown in Tab. \ref{tab:result}.

\begin{table}[h]
\centering
\caption{Evaluation results}\label{tab:result}
\begin{tabular}{ccc}
\hline
Method & $CPC_{ratio}$ & $Value_{ratio}$ \\ \hline
Cost-min 		& 1.0 & 0.362 \\ %\hline
Fb-Control 		& 1.0 & 0.549 \\ %\hline
Fb-Control-M 	& 1.0 & 0.709 \\ %\hline
I-PID 			& 1.0 & 0.892 \\ %\hline
M-PID 			& 1.0 & 0.928 \\ \hline
\end{tabular}
\end{table}

As illustrated in Tab. \ref{tab:result}, all methods could guarantee the CPC constraint, while Cost-min achieves the least advertising value. It is because Cost-min controls CPC greedily and excessively by simply truncating the price, which would lose many valuable ad opportunities. Fb-Control and its modified version, as well as our approaches, also show excellent control capability on achieving the CPC constraint. However, Fb-Control and Fb-Control-M achieved generally lower value than I-PID and M-PID. The key reason is that their generalised bidding strategy is not optimal to address the budget spending and CPC constraint simultaneously. Fb-Control-M obtains a better result than Fb-Control, and justifies our modification. Our approach I-PID outperforms all baseline strategies. M-PID performs even better than I-PID, since the coupling effect is addressed in M-PID and thus the controllers could behave in a more coordinated way. To sum up, compared with the state of the art in the industry practices, our multivariable feedback control systems deliver excellent control capability on the CPC and superior advertising performance in our scenario.

\section{Related Work}
	
	The bid optimization problem is a very actively studied problem in real-time bidding\cite{zhang2012joint, ren2018bidding, maehara2018optimal, zhou2008budget}, and several formulations and algorithms have been proposed in the display advertising scenario. Authors of work  \cite{zhang2014optimal, cai2017real, wu2018budget} proposed models to maximize advertising value within the budget, where the KPI constraint is not considered. Some work has been proposed to specifically address the KPI constraint such as \cite{zhang2016feedback, ghosh2009adaptive}. Little work in diaplay advertising focuses on advertising value and KPI constraints simultaneously. Kitts et al. introduced a generic bidding framework to take into consideration of the advertising value and multiple KPI constraints in \cite{kitts2017ad}. Our work is similar to \cite{kitts2017ad}, however, we focused on the specific KPI constraint and proposed a more delicate strategy. In our work, we abstracted our problem as a linear programming problem and leveraged the primal-dual method. Such an approach is generally applied in the ad allocation scenario \cite{chen2011real, goel2010advertisement, agrawal2014dynamic}. Different from their work, our work adopted this approach to derive the optimal bidding strategy instead of the allocation strategy. To address the dynamic environment, we took advantage of the feedback control theory, which has been proved effective in many scenarios by work of \cite{zhang2016feedback, jambor2012using,  karlsson2013applications}. Other related work includes click-through rate estimation\cite{zhou2018deep, mcmahan2013ad}, conversion rate estimation\cite{yang2017bayesian, lee2012estimating}, winning price prediction \cite{wu2018deep, wang2016functional} and budget pacing\cite{xu2015smart, lee2013real}.

\section{Conclusion}

	In this paper, we focus on the bidding strategy to maximize the advertising value with the budget and the KPI constraint. We convert such a problem into a linear programming problem and leverage the primal-dual method to derive the optimal bidding strategy. The hyper parameters of the bidding strategy is investigated and their relationship with the corresponding constraint is illustrated. It is demonstrated that the hyper parameters have strong control capability on achieving the constraints. Based on our analysis, we propose a feedback control-based solution and design the independent PID control system to address the dynamic environment. To compensate the coupling effect among variables, we further devise the model predictive PID control system by deploying a model predictive module. Extensive experiments are conducted, and our approach is compared with the state of the art in the industry on real-world dataset. The results show that our multivariable control systems deliver superior advertising performance with the KPI constraint holding.
		